\documentclass[10pt,letterpaper,bibnotes,notitlepage,final,balancelastpage,superscriptaddress,twocolumn,showpacs,prl]{revtex4}
\usepackage{amssymb}
\usepackage{amsmath}
\usepackage{amsfonts}
\usepackage{graphicx}

\begin{document}

\title{Robust creation of entanglement between remote memory qubits }
\author{Bo Zhao}
\affiliation{Physikalisches Institut, Universit\"{a}t Heidelberg, Philosophenweg 12,
D-69120 Heidelberg, Germany}
\author{Zeng-Bing Chen}
\affiliation{Hefei National Laboratory for Physical Sciences at Microscale and Department
of Modern Physics, University of Science and Technology of China, Hefei,
Anhui 230026, China}
\author{Yu-Ao Chen}
\affiliation{Physikalisches Institut, Universit\"{a}t Heidelberg, Philosophenweg 12,
D-69120 Heidelberg, Germany}
\author{{J\"{o}rg Schmiedmayer}}
\affiliation{Physikalisches Institut, Universit\"{a}t Heidelberg, Philosophenweg 12,
D-69120 Heidelberg, Germany}
\affiliation{Atominstitut der \"{O}sterreichischen Universit\"{a}ten, TU-Wien, A-1020
Vienna, Austria}
\author{Jian-Wei Pan}
\affiliation{Physikalisches Institut, Universit\"{a}t Heidelberg, Philosophenweg 12,
D-69120 Heidelberg, Germany}
\affiliation{Hefei National Laboratory for Physical Sciences at Microscale and Department
of Modern Physics, University of Science and Technology of China, Hefei,
Anhui 230026, China}

\begin{abstract}
In this Letter we propose a robust quantum repeater architecture
building on the original DLCZ protocol [L.M. Duan \textit{et al.},
Nature \textbf{414}, 413 (2001)]. The architecture is based on
two-photon Hong-Ou-Mandel-type interference which relaxes the long
distance stability requirements by about 7 orders of magnitude,
from sub wavelength for the single photon interference required by
DLCZ to the coherence length of the photons. Our proposal provides
an exciting possibility for robust and realistic long distance
quantum communication.
\end{abstract}

\maketitle

Quantum communication holds the promise in achieving long-distance secure
message transmission by exploiting quantum entanglement between remote
locations \cite{Gisin,Ekert}. For long-distance quantum communication one
must realize quantum network via quantum repeater protocol \cite{repeater},
a combination of entanglement swapping, entanglement purification and
quantum memory. In a seminal paper \cite{DLCZ}, Duan \textit{et al.} (DLCZ)
proposed a promising implementation of the quantum repeater with atomic
ensembles as local memory qubits and linear optics. In the effort of
realizing DLCZ protocol, significant progress has been achieved in recent
years \cite{nonclassical,kuzmich,entanglement}.

However, entanglement generation and entanglement swapping in DLCZ
protocol depend on Mach-Zehnder-type interference. The relative
phase between two remote entangled pairs is sensitive to path
length instabilities, which has to be kept constant within a
fraction of photon's wavelength. Moreover, entanglement generation
and entanglement swapping are probabilistic. If connecting
neighboring entangled pairs doesn't succeed after performing
entanglement swapping, one has to repeat all previous procedures
to reconstruct the entangled pairs. This means the path length
fluctuation must be stabilized until the desired remote entangled
pairs are successfully generated. A particular analysis shows
that, to maintain path length phase instabilities at the level of
$\lambda /10$ ($\lambda$:wavelength; typically $\lambda \sim 1$
$\mu m$ for photons generated from atomic ensembles) requires the
fine control of timing jitter at a sub-femto second level over a
timescale of a few tens of seconds, no matter whether entanglement
generation is performed locally or remotely. It is extremely
difficult for current technology to meet this demanding
requirement, since the lowest reported jitter is about a few tens
of femto-seconds for transferring a timing signal over
kilometer-scale distances for averaging times of $\geq 1$ s
\cite{ye}.

As is well known, the two-photon Hong-Ou-Mandel-type interference
is insensitive to phase instability. The path length fluctuations
should be kept on the length scale within a fraction of photon's
coherence length (say, 1/10 of the coherence length, which is
about 3 m for photons generated from atomic ensembles
\cite{Eisman}). Therefore the robustness is improved about 7
orders of magnitude higher in comparison with the single-photon
Mach-Zehnder-type interference in DLCZ protocol. The interference
of two photons from independent atomic ensembles has been reported
recently \cite{yuan}. This type of two-photon interference has
been widely used in quantum communication and quantum computation \cite%
{two,two-photon}.

To exploit the advantage of two-photon interference, it is natural
to extend the DLCZ protocol by polarization encoding a memory
qubit with two atomic ensembles \cite{kuzmich2}, and entangling
two memory qubits at neighboring sites via a two-photon Bell-state
measurement (BSM). Unfortunately, the BSM won't create the desired
entangled state, but a complex superposition state with spurious
contributions from second-order excitations, which preclude
further entanglement manipulation (see details below).

In this Letter, we explore this problem and find that by appropriate
designing the BSM, the spurious contributions from second-order excitations
can be automatically eliminated when entanglement swapping is performed.
Motivated by this advance we propose a robust quantum repeater architecture
with atomic ensembles and linear optics. This scheme makes use of the
two-photon Hong-Ou-Mandel-type interference, which is about 7 orders of
magnitude more insensitive to path length phase instability than DLCZ
scheme, and thus enables a robust and feasible implementation of
long-distance quantum communication.

The basic element of the quantum repeater is a pencil shaped
atomic sample of $N$ atoms with $\Lambda $ level structure (see
inset in Fig. 1). The write laser pulse induces a spontaneous
Raman process, which prepares the forward-scattered Stokes mode
and collective atomic state into a two mode squeezed state. The
light-atom system can be described as \cite{DLCZ}
\begin{equation}
|\psi \rangle =|0_{a}0_{s}\rangle +\sqrt{\chi }S^{\dagger }a^{\dagger
}|0_{a}0_{s}\rangle +\chi \frac{(S^{\dagger }a^{\dagger })^{2}}{2}%
|0_{a}0_{s}\rangle
\end{equation}%
by neglecting higher order terms, where $|0_{a}\rangle
=\bigotimes_{i}|g\rangle _{i}$ is the ground state of the atomic ensemble
and $|0_{s}\rangle $ denotes vacuum state of the Stokes photon. Here, $%
a^{\dagger }$ is the creation operator of the Stokes mode and the collective
atomic excitation operator is defined by $S^{\dagger }=\frac{1}{\sqrt{N}}%
\sum_{i}|s\rangle _{i}\langle g|$, where $|s\rangle $ is the
metastable atomic state. The small excitation probability $\chi
\ll 1$ can be achieved by manipulating the write laser pulse
\cite{thesis}.
\begin{figure}[tbp]
\begin{center}
\includegraphics[
height=1.8in, width=2.3in ]{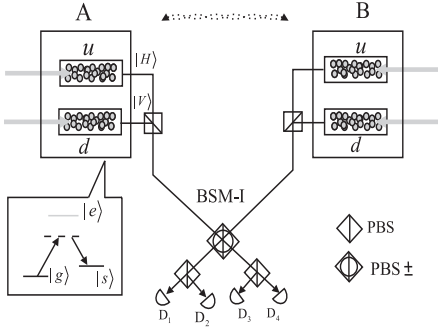}
\end{center}
\caption{Setup for entanglement generation between sites A and B.
Forward-scattered Stokes photons, generated by an off-resonant write laser
pulse via spontaneous Raman transition, are subject to BSM-\uppercase
\expandafter{\romannumeral1} at the middle point. The Stokes photons
generated at the same site are assumed to have different polarization i.e., $%
|H\rangle $ and $|V\rangle $. PBS (PBS$_{\pm }$) reflects photons with
polarization $|V\rangle $ ($|-\rangle $) and transmits photons with
polarization $|H\rangle $ ($|+\rangle $), where $|\pm \rangle =\frac{1}{%
\protect\sqrt{2}}(|H\rangle \pm |V\rangle )$. After passing through the PBS$%
_{\pm }$ and PBS successively, the Stokes photons are detected by single
photon detectors. A coincidence count between single photon detectors D$_{1}$
and D$_{4}$ (D$_{1} $ and D$_{3}$) or D$_{2}$ and D$_{3}$ (D$_{2}$ and D$_{4}
$) will project the four atomic ensembles into the complex entangled state $|%
\protect\psi \rangle _{AB}$ up to a local unitary transformation. The inset
shows the atomic level structure, with the ground state $|g\rangle $,
metastable state $|s\rangle $, and excited state $|e\rangle $.}
\end{figure}

The entanglement generation setup is shown in Fig. 1. Let us consider two
sites A and B at a distance of $L_{0}\leq L_{att}$, with $L_{att}$ the
channel attenuation length. Each site has two atomic ensembles encoded as
one memory qubit and the two atomic ensembles at each node are excited
simultaneously by write laser pulses. We assume the Stokes photons generated
from the two atomic ensembles at the same site have orthogonal polarization
state, e.g., $|H\rangle $ and $|V\rangle $, which denote horizontal and
vertical linear polarization respectively. In this way the memory qubit is
effectively entangled with the polarization state of the emitted Stokes
photon.

The Stokes photons generated from both the sites are directed to the
polarization beam splitter (PBS) and subject to BSM-\uppercase%
\expandafter{\romannumeral1} at the middle point to entangle the two
neighboring memory qubits. However, the two-photon state generated in the
second-order Spontaneous Raman process will also induce a coincidence count
on the detectors. Thus BSM-\uppercase\expandafter{\romannumeral1} can only
prepare the neighboring memory qubits into a complex superposition state
with spurious contributions from second-order excitations. For instance, a
coincidence count between D$_{1}$ and D$_{4}$ projects the two memory qubits
into%
\begin{eqnarray}
|\psi \rangle _{AB} &=&(\frac{S_{u_{A}}^{\dagger }S_{u_{B}}^{\dagger
}+S_{d_{A}}^{\dagger }S_{d_{B}}^{\dagger }}{2}+  \notag \\
&&\frac{S_{u_{A}}^{\dagger 2}+S_{u_{B}}^{\dagger 2}-S_{d_{A}}^{\dagger
2}-S_{d_{B}}^{\dagger 2}}{4})|vac\rangle _{AB}
\end{eqnarray}%
by neglecting high-order terms, where the atomic ensembles are
distinguished by subscript ($u,d$) and ($A,B$). The first part is
the maximally entangled state needed for further operation, while
the second part is the unwanted two-excitation state coming from
second-order excitations. The success probability is on the order
of $O(\chi ^{2}\eta ^{2}e^{-L_{0}/L_{att}})$ by considering the
channel attenuation, where $\eta $ is the detection
efficiency. The time needed in this process is $T_{0}\approx \frac{T_{cc}}{%
\chi ^{2}\eta ^{2}e^{-L_{0}/L_{att}}}$, with $T_{cc}=L_{0}/c$ the classical
communication time.

It is obvious the spurious contributions of two-excitation terms
prevent further entanglement manipulation and must be eliminated
by some means. However we find that it is not necessary to worry
about these terms, because they can be automatically washed out if
the BSM in entanglement swapping is carefully designed. In ideal
case a maximally entangled state can be created by implementing
entanglement swapping.

\begin{figure}[ptb]
\begin{center}
\includegraphics[
height=1.8in, width=2.3in]{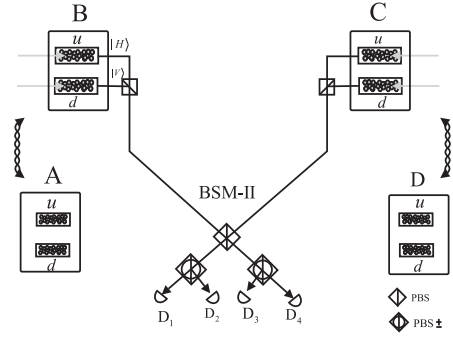}
\end{center}
\caption{Setup for entanglement connection between sites A and D
via entanglement swapping. Complex entangled states have been
prepared in the memory qubits between sites (A,B) and (C,D). The
memory qubits at site B and C are illuminated by near resonant
read laser pulses, and the retrieved anti-Stokes photons are
subject to BSM-\uppercase\expandafter{\romannumeral2} at the
middle point. The anti-Stokes photons at the same site have
different
polarizations $|H\rangle$ and $|V\rangle$. After passing through PBS and PBS$%
_{\pm}$ successively, the anti-Stokes photons are detected by single photon
detectors. Coincidence count between D$_{1}$ and D$_{4}$ (D$_{1}$ and D$_{3}$%
) or D$_{2}$ and D$_{3}$ (D$_{2}$ and D$_{4}$) are registered. The memory
qubits will be projected into an effectively maximally entangled state $%
\protect\rho_{AD}$ up to a local unitary transformation. Note that
the PBS  arrangement in BSM-\uppercase\expandafter{\romannumeral2}
and BSM-\uppercase\expandafter{\romannumeral1} is different.}
\end{figure}

The entanglement swapping setup is depicted in Fig. 2. Let us
consider four communication sites (A,B) and (C,D) and assume we
have created complex entangled states $|\psi \rangle _{AB}$ and
$|\psi \rangle _{CD}$ between sites (A,B) and (C,D) respectively.
The memory qubits at site B and C are illuminated simultaneously
by read laser pulses. The retrieved anti-Stokes photons are
subject to BSM-\uppercase\expandafter{\romannumeral2} at the
middle point between B and C (see Fig. 2). Note that the
arrangement of the PBSs in
BSM-\uppercase\expandafter{\romannumeral2} is carefully designed,
so that the two-photon states converted from the unwanted
two-excitation terms are directed into the same output and thus
won't induce a coincidence count on the detectors. In ideal case,
if the retrieve efficiency is unit and perfect photon detectors
are used to distinguish photons' numbers, only the two-photon
coincidence count will be registered and project the memory qubits
into a maximally entangled state. For instance, when a coincidence
count between D$_{1}$ and D$_{4}$ is registered
one will obtain%
\begin{equation}
|\phi ^{+}\rangle _{AD}=(S_{u_{A}}^{\dagger }S_{u_{D}}^{\dagger
}+S_{d_{A}}^{\dagger }S_{d_{D}}^{\dagger })/\sqrt{2}|vac\rangle
_{AD}.
\end{equation}%
In this way a maximally entangled state across sites A and D is
generated by performing entanglement swapping.

However, for realistic atomic ensembles the retrieve efficiency
$\eta _{r}$ is determined by optical depth of the atomic ensemble
\cite{optimal}, and current single photon detectors are incapable
of distinguishing photon numbers. Taking into account these
imperfections, the multi-photon coincidence counts in
BSM-\uppercase\expandafter{\romannumeral2} have to be considered.
Through some simple calculations, one can find that the
coincidence counts will prepare the memory qubits into a mixed
entangled
state of the form%
\begin{equation}
\rho _{AD}=p_{2}\rho _{2}+p_{1}\rho _{1}+p_{0}\rho _{0},
\end{equation}%
where the coefficients $p_{2}$, $p_{1}$ and $p_{0}$ are determined by the
retrieve efficiency and detection efficiency \cite{zhao}. Here $\rho
_{2}=|\phi ^{+}\rangle _{AD}\langle \phi ^{+}|$ is a maximally entangled
state, $\rho _{1}$ is a maximally mixed state where only one of the four
atomic ensembles has one excitation and $\rho _{0}$ is the vacuum state that
all the atomic ensembles are in the ground states.

\begin{figure}[ptb]
\begin{center}
\includegraphics[
height=1.8in, width=2.3in ]{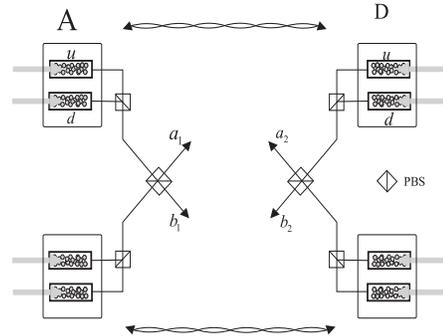}
\end{center}
\caption{Setup for quantum entanglement purification. Effectively entangled
states have been prepared in the memory qubits between two distant sites A
and D. The memory qubits at the two sites are illuminated by near resonant
read laser pulse, and the retrieved entangled photon pairs are directed to
two PBS respectively. The photons in mode $b_{1}$ and $b_{2}$ are detected
in $|\pm \rangle =\frac{1}{\protect\sqrt{2}}(|H\rangle \pm |V\rangle )$
basis and the left photons in mode $a_{1}$ and $a_{2}$ are restored in the
memory qubits at the two sites respectively.}
\end{figure}

It is easy to see that $\rho _{AD}$ is in fact an effectively maximally
entangled states, which can be projected automatically to a maximally
entangled state in the entanglement based quantum cryptography schemes. When
we implement quantum cryptography via Ekert protocol \cite{Ekert}, only the
first term $\rho _{2}$ can contribute to a coincidence count between the
detectors at the two sites and will be registered after classical
communication. The maximally mixed state term $\rho _{1}$ and the vacuum
term $\rho _{0}$ have no contribution to the experimental results, and thus $%
\rho _{AD}$ is equivalent to the Bell state $|\phi ^{+}\rangle
_{AD}=(S_{u_{A}}^{\dagger }S_{u_{D}}^{\dagger }+S_{d_{A}}^{\dagger
}S_{d_{D}}^{\dagger })/\sqrt{2}|vac\rangle _{AD}$.

The effectively entangled state can be connected to longer communication
distance via further entanglement swapping. Taking into account higher-order
excitations, the effectively mixed entangled pair reads $\rho ^{\prime
}=\rho +p_{h}^{\prime }\rho _{h}^{\prime }$, where the normalized mixed
state $\rho _{h}^{\prime }$ denotes contributions from higher-order
excitations and the small coefficients $p_{h}^{\prime }$ is on the order of $%
O(\chi )\ll 1$ \cite{zhao}. After the \textit{j}-th ($j\geq 2$)
swapping step, the effective entangled pair can be described as
\begin{equation}
\rho _{s_{j}}^{\prime }=p_{2s_{j}}\rho _{2s_{j}}+p_{1s_{j}}\rho
_{1s_{j}}+p_{0s_{j}}\rho _{0s_{j}}+p_{hs_{j}}^{\prime }\rho
_{hs_{j}}^{\prime }.
\end{equation}%
Here $\rho _{2s_{j}}$ is the maximally entangled state between two
memory qubits at a distance of $L=(2^{j+1}-1)L_{0}$, and $\rho
_{1s_{j}},$ $\rho _{0s_{j}}$ are also the maximally mixed state
and vacuum state respectively. Note that $\rho _{s_{1}}^{\prime
}=\rho ^{\prime }$ is just the mixed entangled state created after
the first entanglement swapping step. The coefficients can be
estimated to be \cite{zhao}
\begin{eqnarray}
p_{hs_{j}}^{\prime } &\sim &O(j\chi ) \\
p_{\alpha s_{j}} &\approx &p_{\alpha s_{j-1}}+O(j\chi ),(\alpha =0,1,2)
\end{eqnarray}%
It is readily to find that the contributions from higher-order excitations
can be safely neglected, as long as the small excitation probability
fulfills $j\chi \ll 1$, which can be easily achieved by tuning the write
laser pulse. One can also see that the coefficients $p_{2s_{j}}$, $%
p_{1s_{j}} $ and $p_{0s_{j}}$ are stable to the first order, therefore the
probability to find an entangled pair in the remaining memory qubits is
almost a constant and won't decrease significantly with distance during the
entanglement connection process. The time needed for the \textit{j-}th
connection step satisfies the iteration formula $T_{s_{j}}=\frac{1}{p_{s_{j}}%
}[T_{s_{j-1}}+2^{j}T_{cc}],$ where the success probability $p_{s_{j}}$ is on
the order of $O(\eta _{r}^{2}\eta ^{2}e^{-L_{0}/L_{att}})$. The total time
needed for the entanglement connection process is%
\begin{equation}
T_{tot}\approx T_{0}\prod\limits_{j}p_{s_{j}}^{-1}\approx \frac{T_{cc}}{\chi
^{2}}e^{L_{0}/Latt}(L/L_{0})^{\log _{2}^{1/p}},
\end{equation}%
where $p=\eta _{r}^{2}\eta ^{2}e^{-L_{0}/L_{att}}$ is a constant. The
excitation probability can be estimated to be $\chi \sim L_{0}/L$, and then
the time needed in the entanglement connection process $T_{tot}\propto $ $%
(L/L_{0})^{2+\log _{2}^{1/p}}$ scales polynomially or quadratically with the
communication distance.

The effectively maximally entangled state generated above may be
imperfect due to decoherence. For simplicity, assume a mixed state
$\rho _{2}=F|\phi ^{+}\rangle _{AD}\langle \phi ^{+}|+(1-F)|\psi
^{+}\rangle _{AD}\langle \psi ^{+}|$ is created after entanglement
swapping, where $|\psi ^{+}\rangle _{AD}=(S_{u_{A}}^{\dagger
}S_{d_{D}}^{\dagger }+S_{d_{A}}^{\dagger }S_{u_{D}}^{\dagger
})/\sqrt{2}|vac\rangle _{AD}$. The mixed entangled state
can be purified by linear-optics entanglement purification protocol \cite%
{purification}. As shown in Fig. 3, two effectively mixed
entangled pairs are created in parallel via entanglement swapping.
The effectively entangled states stored in the four memory qubits
are converted into entangled photons by the read laser pulses, and
then subject to two PBSs respectively. The photons in mode $b_{1}$
and $b_{2}$ are detected in $|\pm \rangle
=\frac{1}{\sqrt{2}}(|H\rangle \pm |V\rangle )$ basis by single
photon detectors, and will project the photons in mode $a_{1}$ and
$a_{2}$ into an effectively maximally entangled state of higher
fidelity $F^{\prime }=\frac{F^{2}}{F^{2}+(1-F)^{2}}$
\cite{purification}. The higher-fidelity entangled pair in mode
$a_{1}$ and $a_{2}$ can be restored into two distant memory qubits
by means of dark-state-polariton \cite{memory1} for further
manipulation.

To generate a remote entangled pair, nested quantum purification
has to be implemented. The total time overhead to create
entanglement across two communication nodes at a distance of 1270
km can be numerically estimated. In our calculation, we assume the
distance $L_{0}=10$ km and thus the connection step $j=6$. The
photon loss rate is considered to be $0.1$ dB/km in free space as
reported \cite{peng}. The initial fidelity is assumed to be
$F=0.88$. To increase the efficiency, we assume high efficiency
(99\%) photon counting detectors based on atomic ensembles are
used \cite{kwait}, and the retrieve efficiency is considered to be
$98\%$. Entanglement purification is performed twice during the
whole process to improve the fidelity. Our numerical results give
a total time of about three hours to create an effectively
entangled pair, with a probability of $0.85$ to get the entangled
pair of fidelity $95\%$ \cite{zhao}. We note that the time
overhead can be reduced significantly by optimization.

In our scheme, entanglement creation and swapping are both
performed remotely, since they rely on two-photon interference.
For the sake of scalability, entanglement generation could be
locally performed, because it is usually rate-limiting stage due
to the low excitation probability. The prize to pay is that one
has to manipulate at least two memory qubits at each site. The
locally entangled pair can also be generated by storing
polarization entangled photons. Let us assume memory qubits A and
B (see Fig. 1) are at one site and that we have created
polarization entangled photon pair with the help of single photon
source \cite{event}. By sending the two entangled photons into
memory qubits A and B respectively and storing them via
dark-state-polariton, one can generate local entanglement between
memory qubits A and B. Entanglement swapping and entanglement
purification discussed above also apply to the entangled memory
qubits and thus allow implementation of a robust quantum repeater.
In this case, on demand single photon source will help to improve
the scalability significantly \cite{kuzmich}.

In summary, we have proposed a robust and feasible quantum repeater by means
of two photon Hong-Ou-Mandel-type interference. Long-lived quantum memory is
crucial for implementing the atomic ensemble based quantum repeater. In a
recent proposal \cite{hours}, it was shown that the ground state of $^{3}$He
has the potential to store a quantum state for times as long as hours.

This work was supported by the DFG, the NFRP and the European
Commission.

\textit{Note added.}--- After the paper was finished, a similar
work by L. Jiang \textit{et al.} \cite{jiang} appeared on arXiv.

\end{document}